\documentclass[aps,preprint,showpacs,psfig]{revtex4}
\usepackage{amsmath}
\usepackage{amssymb}
\usepackage{graphicx}
\usepackage{dcolumn}
\usepackage{bm}
\usepackage{psfig}

\input{tcilatex}

\begin{document}

\title{Spin Current and Current-Induced Spin Transfer Torque in
Ferromagnet-Quantum Dot-Ferromagnet Coupled Systems}
\author{Hai-Feng Mu, Gang Su$^{\ast }$, and Qing-Rong Zheng}
\affiliation{College of Physical Sciences, Graduate University of Chinese Academy of
Sciences, P.O. Box 4588, Beijing 100049, China}

\begin{abstract}
Based on Keldysh's nonequilibrium Green function method, the spin-dependent
transport properties in a ferromagnet-quantum dot (QD)-ferromagnet coupled
system are investigated. It is shown the spin current shows quite different
characteristics from its electrical counterpart, and by changing the
relative orientation of both magnetizations, it can change its magnitude
even sign. The current-induced spin transfer torque (CISTT) is uncovered to
be greatly enhanced when the bias voltage meets with the discrete levels of
the QD at resonant positions. The relationship between the CISTT, the
electrical current and the spin current is also addressed.
\end{abstract}

\maketitle

\section{Introduction}

Investigations on the spin-dependent transport in a magnetic tunnel junction
(MTJ) have attracted much interest in the last decade, as the MTJs have
essential applications in spintronic devices\cite%
{Prinz,Wolf,Zutic,Maekawa,su}. One of phenomena in MTJs is the so-called
tunnel magnetoresistance (TMR) effect, which states that the tunneling
current through the junction depends sensitively on the relative orientation
of the magnetizations of both ferromagnetic electrodes, that is caused by
the spin-dependent scattering of conduction electrons. The difference
between the currents through spin up and down channels in MTJs is usually
referred to as the spin current\cite{Maekawa}. On the other hand, a reverse
effect to TMR was predicted independently by Slonczewski\cite{Slonczewski}
and Berger\cite{Berger}, i.e., the spin-polarized electrons passing from the
left ferromagnetic layer into the right layer in which the magnetization
deviates the left by an angle may exert a torque to the right ferromagnet
and can change the orientation of its magnetization. This effect was coined
as the spin transfer effect (for review, see e.g. Ref.\cite{L.Berger,su}),
which can lead to the current-induced magnetization reversal, and might
offer a promise for the current-controlled spintronic devices. It has
recently attracted intensive investigations both experimentally\cite%
{Tsoi,Myers,Katine,Wegrowe,Sun,Asamitsu} and theoretically\cite%
{Waintal,Brataas,Zhu,Tserkovnyak,X.Waintal,Brouwer,Mu,Parcollet}.

Among other magnetic tunnel structures (e.g. Refs.\cite{mu0}), the
ferromagnet-quantum dot-ferromagnet (FM-QD-FM) coupled systems have also
received much attention recently. Previous theoretical works on
spin-dependent transport through QDs are mainly focused on the tunnel
electrical current and the TMR effect for collinear configuration\cite%
{Barnas,Fransson,Zhang,Bulka,Lopez,Martinek,Pedersen} and noncollinear
configuration\cite%
{Sergueev,Konig,zhu1,Braun,Rudzinski,Wetzels,Fran1,Fran2,Weymann}. However,
the investigations on the spin current and the current-induced spin transfer
torque (CISTT) in such systems are sparsely reported. It is the purpose of
this paper to study the spin current and spin transfer torque in FM-QD-FM
coupled MTJs.

In terms of Keldysh's nonequilibrium Green function method, the tunnel
electrical current, the spin current and the spin transfer torque in the
FM-QD-FM coupled system will be investigated. It is found the spin current
exhibits different behaviors from its electrical counterpart. It is also
shown that the resonant positions for the tunneling electrical current and
spin current become far separated with the increase of the Coulomb
interaction $U$ in the QD. The magnitudes of the CISTT at two resonant
positions are found to be greatly enhanced at resonant positions. At the
bias larger than $(\varepsilon _{0}+U)/e$, the CISTT can reach a saturation
plateau which is independent of the Coulomb interaction. The CISTT shows a
kink-like behavior with the increase of the spin current. The relationship
between the CISTT, the electrical current and the spin current is also
addressed.

The rest of this paper is outlined as follows. In Sec. II, the model and
formalism for the tunnel electrical current, the spin current and the
current-induced spin transfer torque will be established. In Sec. III, the
corresponding numerical results will be given. Finally, a summary will be
presented.

\section{Model and Formalism}

Let us consider a single-level QD coupled to two ferromagnetic electrodes.
The left (L) and right (R) electrodes are connected with the bias voltage $%
V/2$ and $-V/2$, respectively, as shown in Fig. 1. The magnetic moment $%
\mathbf{M}_{L}$ of the left FM is assumed to be parallel to the $z$ axis,
while the moment $\mathbf{M}_{R}$ of the right FM is aligned along the $%
z^{\prime }$ axis which deviates the $z$ axis by a relative angle $\theta $.
The tunnel current flows along the $x$ axis and perpendicular to the
junction plane. The system consists of the left, right, QD and coupling
parts, and can be described by the following Hamiltonian:

\begin{equation}
H=H_{L}+H_{R}+H_{d}+H_{T},
\end{equation}%
with 
\begin{equation}
H_{L}=\underset{k\sigma }{\sum }\varepsilon _{kL\sigma }a_{kL\sigma }^{\dag
}a_{kL\sigma },
\end{equation}

\begin{equation}
H_{R}=\underset{k\sigma }{\sum }[\varepsilon _{R}(k)-\sigma \mathbf{M}%
_{R}\cos \theta ]a_{kR\sigma }^{\dag }a_{kR\sigma }-\mathbf{M}_{R}\sin
\theta a_{kR\sigma }^{\dag }a_{kR\bar{\sigma}},
\end{equation}

\begin{equation}
H_{d}=\underset{\sigma }{\sum }\varepsilon _{0}c_{\sigma }^{\dag }c_{\sigma
}+Un_{\uparrow }n_{\downarrow },
\end{equation}

\begin{equation}
H_{T}=\underset{k\alpha \sigma }{\sum }T_{k\alpha }a_{k\alpha \sigma }^{\dag
}c_{\sigma }+h.c.,\text{ }\alpha =L,R.
\end{equation}%
where $\varepsilon _{k\alpha \sigma }=\varepsilon _{\alpha }(k)-\sigma 
\mathbf{M}_{\alpha }-eV_{\alpha }$ is the single-electron energy for the
wavevector $\mathbf{k}$ and spin $\sigma $ in the $\alpha $ electrode, $%
a_{k\alpha \sigma }$ and $c_{\sigma }$ are annihilation operators of
electrons with spin $\sigma $ in the $\alpha $ electrode and the QD,
respectively, $n_{\sigma }=c_{\sigma }^{\dag }c_{\sigma }$, $U$ represents
the on-site Coulomb interaction between electrons in the QD, and $T_{k\alpha
}$ is the coupling matrix elements between the $\alpha $ electrode and the
QD.

\begin{figure}[tbp]
\vspace{-1.5cm}\includegraphics[width=0.85\linewidth,clip]{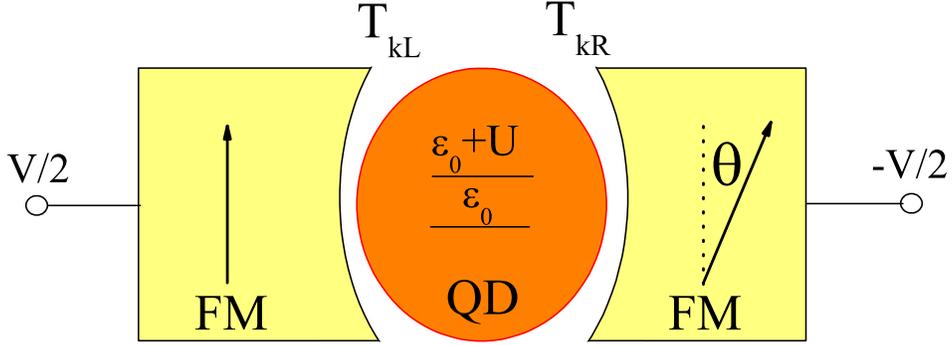} \vspace{%
-1.5cm}
\caption{Schematic illustration of the system consisting of two ferromagnets
and a quantum dot separated by the tunnel barriers, where $T_{k\protect%
\alpha }$ ($\protect\alpha =L,R$) stands for the coupling matrix between the 
$\protect\alpha $ electrode and the QD, and both magnetizations are aligned
by a relative angle $\protect\theta $.}
\label{Fig.1}
\end{figure}

\subsection{Tunnel electrical current and spin current}

The tunnel electrical current is composed of the sum of the currents carried
by spin up and down electrons:%
\begin{equation}
I(V)=I_{L\uparrow }(V)+I_{L\downarrow }(V),
\end{equation}%
while the spin current is defined by the difference between the electrical
currents through the spin up and down channels\cite{Maekawa}: 
\begin{equation}
I_{s}(V)=I_{L\uparrow }(V)-I_{L\downarrow }(V),
\end{equation}%
where%
\begin{eqnarray*}
I_{L\uparrow }(V) &=&-\frac{2e}{\hbar }\Re e\sum_{kL}T_{kL}G_{kL}^{\uparrow
\uparrow ,<}(t,t), \\
I_{L\downarrow }(V) &=&-\frac{2e}{\hbar }\Re
e\sum_{kL}T_{kL}G_{kL}^{\downarrow \downarrow ,<}(t,t).
\end{eqnarray*}%
with $G_{kL}^{\sigma ^{\prime }\sigma ,<}(t,t^{\prime })=i\langle
a_{kL\sigma }^{\dag }(t^{\prime })c_{\sigma ^{\prime }}(t)\rangle $ the
lesser Green function.

By applying the Langrenth theorem and Fourier transform, we may obtain the
following equation%
\begin{equation}
G_{kL}^{<}(\varepsilon )=T_{kL}^{\ast }[G^{r}(\varepsilon
)g_{kL}^{<}(\varepsilon )+G^{<}(\varepsilon )g_{kL}^{a}(\varepsilon )],
\label{sel-L}
\end{equation}%
where $G_{k\alpha }^{r,<}(\varepsilon )=\left( 
\begin{array}{cc}
G_{k\alpha }^{\uparrow \uparrow r,<}(\varepsilon ) & G_{k\alpha }^{\uparrow
\downarrow r,<}(\varepsilon ) \\ 
G_{k\alpha }^{\downarrow \uparrow r,<}(\varepsilon ) & G_{k\alpha
}^{\downarrow \downarrow r,<}(\varepsilon )%
\end{array}%
\right) ,$ $G^{r,<}(\varepsilon )=\left( 
\begin{array}{cc}
G^{\uparrow \uparrow r,<}(\varepsilon ) & G^{\uparrow \downarrow
r,<}(\varepsilon ) \\ 
G^{\downarrow \uparrow r,<}(\varepsilon ) & G^{\downarrow \downarrow
r,<}(\varepsilon )%
\end{array}%
\right) ,$ $g_{k\alpha }^{<}(\varepsilon )=\left( 
\begin{array}{cc}
i2\pi f_{\alpha }(\varepsilon _{k\alpha \uparrow })\delta (\varepsilon
-\varepsilon _{k\alpha \uparrow }) &  \\ 
& i2\pi f_{\alpha }(\varepsilon _{k\alpha \downarrow })\delta (\varepsilon
-\varepsilon _{k\alpha \downarrow })%
\end{array}%
\right) $, $g_{k\alpha }^{r,a}(\varepsilon )=\left( 
\begin{array}{cc}
\frac{1}{\varepsilon -\varepsilon _{k\alpha \uparrow }\pm i\eta } &  \\ 
& \frac{1}{\varepsilon -\varepsilon _{k\alpha \downarrow }\pm i\eta }%
\end{array}%
\right) $, $G^{r}(\varepsilon )$ and $G^{<}(\varepsilon )$ are the retarded
and lesser Green functions of the QD, respectively, and $f_{\alpha
}(\varepsilon )$ is the Fermi distribution function in the $\alpha $
electrode. By defining $M(\varepsilon )\mathbf{=[}f_{L}(\varepsilon )\mathbf{%
(}G^{r}(\varepsilon )-G^{a}(\varepsilon ))+G^{<}(\varepsilon )]\Gamma
_{L}(\varepsilon )$, we can simplify $I_{\uparrow (\downarrow )}(V)$ as 
\begin{equation*}
I_{\uparrow (\downarrow )}(V)=-\frac{ie}{\hbar }\int \frac{d\varepsilon }{%
2\pi }M_{\uparrow \uparrow (\downarrow \downarrow )}(\varepsilon ),
\end{equation*}%
where $M=\left( 
\begin{array}{cc}
M_{\uparrow \uparrow } & M_{\uparrow \downarrow } \\ 
M_{\downarrow \uparrow } & M_{\downarrow \downarrow }%
\end{array}%
\right) $ is a $2\times 2$ matrix, $\Gamma _{\alpha }(\varepsilon )=\left( 
\begin{array}{cc}
\Gamma _{\alpha \uparrow }(\varepsilon ) &  \\ 
& \Gamma _{\alpha \downarrow }(\varepsilon )%
\end{array}%
\right) $ with $\Gamma _{\alpha \sigma }(\varepsilon )=2\pi \underset{%
k\alpha }{\sum }|T_{k\alpha }|^{2}\delta (\varepsilon -\varepsilon _{k\alpha
\sigma })$.

The lesser Green function $G^{<}$ can be calculated by the Keldysh equation $%
G^{<}=G^{r}\Sigma ^{<}G^{a}$. To get $\Sigma ^{<}$, we invoke Ng's ansatz%
\cite{Ng}: $\Sigma ^{<}=\Sigma _{0}^{<}B$, where $\Sigma
_{0}^{<}(\varepsilon )=i[f_{L}(\varepsilon )\Gamma _{L}+f_{R}(\varepsilon )%
\mathcal{R}\Gamma _{R}\mathcal{R}^{\dag }]$, $B=(\Sigma _{0}^{r}-\Sigma
_{0}^{a})^{-1}(\Sigma ^{r}-\Sigma ^{a})$, $\Sigma _{0}^{r}(\varepsilon
)-\Sigma _{0}^{a}(\varepsilon )=-i[\Gamma _{L}+\mathcal{R}\Gamma _{R}%
\mathcal{R}^{\dag }]$, $\Sigma ^{r}(\varepsilon )-\Sigma ^{a}(\varepsilon
)=G^{a-1}-G^{r-1}$, and $\mathcal{R}=\left( 
\begin{array}{cc}
\cos \frac{\theta }{2} & -\sin \frac{\theta }{2} \\ 
\sin \frac{\theta }{2} & \cos \frac{\theta }{2}%
\end{array}%
\right) $. Under these considerations, we finally get 
\begin{equation}
I_{\uparrow (\downarrow )}(V)=\frac{e}{\hbar }\int \frac{d\varepsilon }{2\pi 
}(f_{R}-f_{L})X_{\uparrow \uparrow (\downarrow \downarrow )}.
\end{equation}%
where $X=G^{r}\mathcal{R}\Gamma _{R}\mathcal{R}^{\dag }BG^{a}\Gamma _{L}$ $%
=\left( 
\begin{array}{cc}
X_{\uparrow \uparrow } & X_{\uparrow \downarrow } \\ 
X_{\downarrow \uparrow } & X_{\downarrow \downarrow }%
\end{array}%
\right) $ is also a $2\times 2$ matrix.

Consequently, the tunnel current and spin current have the forms of%
\begin{eqnarray}
I(V) &=&\frac{e}{\hbar }\int \frac{d\varepsilon }{2\pi }(f_{R}-f_{L})TrX, \\
I_{s}(V) &=&\frac{e}{\hbar }\int \frac{d\varepsilon }{2\pi }(f_{R}-f_{L})Tr(X%
\hat{\sigma}_{3}).
\end{eqnarray}%
with $\hat{\sigma}_{3}=\left( 
\begin{array}{cc}
1 & 0 \\ 
0 & -1%
\end{array}%
\right) $ the Pauli matrix.

\subsection{Current-induced spin transfer torque}

The spin torque exerting on the right ferromagnet is determined by the time
evolution rate of the total spin in the right ferromagnet\cite%
{Slonczewski,Berger}, which is composed of two parts: one is caused by the
spin-dependent potential that is known as the equilibrium torque, the other
is the current-induced spin transfer torque caused by the tunnel Hamiltonian 
$H_{T}$. By means of the nonequilibrium Green functions, the CISTT exerting
on the right ferromagnet can be obtained by\cite{Zhu} 
\begin{equation}
\tau ^{Rx^{\prime }}=-\cos \theta \Re e\underset{kR}{\sum }\int \frac{%
d\varepsilon }{2\pi }Tr[G_{kR}^{<}(\varepsilon )\hat{\sigma}_{1}T_{kR}^{\ast
}]+\sin \theta \Re e\underset{kR}{\sum }\int \frac{d\varepsilon }{2\pi }%
Tr[G_{kR}^{<}(\varepsilon )\hat{\sigma}_{3}T_{kR}^{\ast }],  \label{torque}
\end{equation}%
where $\hat{\sigma}_{1}=\left( 
\begin{array}{cc}
0 & 1 \\ 
1 & 0%
\end{array}%
\right) $.

By a treatment similar to Eq. (\ref{sel-L}), we can obtain 
\begin{equation*}
G_{kR}^{<}(\varepsilon )=T_{kR}[G^{r}(\varepsilon )\mathcal{R}%
g_{kR}^{<}(\varepsilon )\mathcal{R}^{\dag }+G^{<}(\varepsilon )\mathcal{R}%
g_{kR}^{a}(\varepsilon )\mathcal{R}^{\dag }].
\end{equation*}%
Therefore, the CISTT can be rewritten as 
\begin{equation}
\tau ^{Rx^{\prime }}=\frac{1}{4\pi }\int d\varepsilon
(f_{R}-f_{L})Tr[G^{r}(\varepsilon )\Gamma _{L}BG^{a}(\varepsilon )\mathcal{R}%
\Gamma _{R}\mathcal{R}^{\dag }(-\cos \theta \hat{\sigma}_{1}+\sin \theta 
\hat{\sigma}_{3})].
\end{equation}%
The remaining task is to calculate the retarded Green function $G^{r}$.

By the equation of motion, we can derive 
\begin{equation}
(\varepsilon -\varepsilon _{0})G^{r}(\varepsilon )=\mathbf{1}+\underset{%
k\alpha }{\sum }T_{k\alpha }^{\ast }G_{k\alpha }^{r}(\varepsilon
)+UG^{(2)r}(\varepsilon ),  \label{eqn-motion}
\end{equation}%
where $\mathbf{1}$ is a unit matrix, $G^{(2)r}(\varepsilon )=\left( 
\begin{array}{cc}
G^{\uparrow \uparrow (2)r}(\varepsilon ) & G^{\uparrow \downarrow
(2)r}(\varepsilon ) \\ 
G^{\downarrow \uparrow (2)r}(\varepsilon ) & G^{\downarrow \downarrow
(2)r}(\varepsilon )%
\end{array}%
\right) ,$ $G_{k\alpha }^{\sigma ^{\prime }\sigma ,r}(t-t^{\prime
})=-i\theta (t-t^{\prime })\langle \{a_{k\alpha \sigma ^{\prime
}}(t),c_{\sigma }^{\dag }(t^{\prime })\}\rangle ,$ $G^{\sigma ^{\prime
}\sigma ,(2)r}(t-t^{\prime })=-i\theta (t-t^{\prime })\langle \{c_{\sigma
^{\prime }}(t)n_{\bar{\sigma}^{\prime }}(t),c_{\sigma }^{\dag }(t^{\prime
})\}\rangle ,$ and 
\begin{eqnarray*}
G_{kL}^{r} &=&T_{kL}g_{kL}^{r}G^{r}(\varepsilon ), \\
G_{kR}^{r} &=&T_{kR}\mathcal{R}g_{kR}^{r}\mathcal{R}^{\dag
}G^{r}(\varepsilon ).
\end{eqnarray*}%
Substituting these equations into Eq. (\ref{eqn-motion}), up to the
third-order of Green functions, we arrive at 
\begin{equation}
(\varepsilon -\varepsilon _{0}-U)G^{(2)r}(\varepsilon )=N+\underset{k\alpha }%
{\sum }[T_{k\alpha }^{\ast }G_{k\alpha }^{1(2)r}(\varepsilon )+T_{k\alpha
}G_{k\alpha }^{2(2)r}(\varepsilon )-T_{k\alpha }^{\ast }G_{k\alpha
}^{3(2)r}(\varepsilon )],  \label{Self-con}
\end{equation}%
where $N=\left( 
\begin{array}{cc}
\langle n_{\downarrow }\rangle & -\langle c_{\downarrow }^{\dag }c_{\uparrow
}\rangle \\ 
-\langle c_{\uparrow }^{\dag }c_{\downarrow }\rangle & \langle n_{\uparrow
}\rangle%
\end{array}%
\right) ,$ $G_{k\alpha }^{i(2)r}=\left( 
\begin{array}{cc}
G_{k\alpha }^{\uparrow \uparrow i(2)r} & G_{k\alpha }^{\uparrow \downarrow
i(2)r} \\ 
G_{k\alpha }^{\downarrow \uparrow i(2)r} & G_{k\alpha }^{\downarrow
\downarrow i(2)r}%
\end{array}%
\right) $ $(i=1,2,3)$,%
\begin{eqnarray*}
G_{k\alpha }^{\sigma ^{\prime }\sigma ,1(2)r}(t-t^{\prime }) &=&-i\theta
(t-t^{\prime })\langle \{a_{k\alpha \sigma ^{\prime }}(t)n_{\bar{\sigma}%
^{\prime }}(t),c_{\sigma }^{\dag }(t^{\prime })\}\rangle , \\
G_{k\alpha }^{\sigma ^{\prime }\sigma ,2(2)r}(t-t^{\prime }) &=&-i\theta
(t-t^{\prime })\langle \{a_{k\alpha \bar{\sigma}^{\prime }}^{\dag
}(t)c_{\sigma ^{\prime }}(t)c_{\bar{\sigma}^{\prime }}(t),c_{\sigma }^{\dag
}(t^{\prime })\}\rangle , \\
G_{k\alpha }^{\sigma ^{\prime }\sigma ,3(2)r}(t-t^{\prime }) &=&-i\theta
(t-t^{\prime })\langle \{a_{k\alpha \bar{\sigma}^{\prime }}(t)c_{\bar{\sigma}%
^{\prime }}^{\dag }(t)c_{\sigma ^{\prime }}(t),c_{\sigma }^{\dag }(t^{\prime
})\}\rangle .
\end{eqnarray*}%
\begin{eqnarray*}
g_{kL}^{r-1}G_{kL}^{1(2)r}(\varepsilon ) &=&T_{kL}G^{(2)r}(\varepsilon )+%
\underset{k^{\prime }\alpha }{\sum }[-T_{k^{\prime }\alpha }G_{kLk^{\prime
}\alpha }^{1(3)r}(\varepsilon )+T_{k^{\prime }\alpha }^{\ast }G_{kLk^{\prime
}\alpha }^{1(3)r}(\varepsilon )], \\
\tilde{g}_{kL}^{r-1}G_{kL}^{2(2)r}(\varepsilon ) &=&T_{kL}^{\ast
}G^{(2)r}(\varepsilon )+\underset{k^{\prime }\alpha }{\sum }T_{k^{\prime
}\alpha }^{\ast }[G_{kLk^{\prime }\alpha }^{3(3)r}(\varepsilon
)+G_{kLk^{\prime }\alpha }^{4(3)r}(\varepsilon )], \\
g_{\bar{k}L}^{r-1}G_{kL}^{3(2)r}(\varepsilon ) &=&T_{kL}[G^{r}(\varepsilon
)-G^{(2)r}(\varepsilon )]+\underset{k^{\prime }\alpha }{\sum }[-T_{k^{\prime
}\alpha }G_{kLk^{\prime }\alpha }^{5(3)r}(\varepsilon )+T_{k^{\prime }\alpha
}^{\ast }G_{kLk^{\prime }\alpha }^{6(3)r}(\varepsilon )],
\end{eqnarray*}%
with $\tilde{g}_{k\alpha }^{r}(\varepsilon )=\left( 
\begin{array}{cc}
\tilde{g}_{k\alpha }^{\uparrow r}(\varepsilon ) &  \\ 
& \tilde{g}_{k\alpha }^{\downarrow r}(\varepsilon )%
\end{array}%
\right) =\left( 
\begin{array}{cc}
\frac{1}{\varepsilon -2\varepsilon _{0}-U+\varepsilon _{k\alpha \downarrow
}+i\eta } &  \\ 
& \frac{1}{\varepsilon -2\varepsilon _{0}-U+\varepsilon _{k\alpha \uparrow
}+i\eta }%
\end{array}%
\right) ,$ $g_{\bar{k}\alpha }^{r}(\varepsilon )=\left( 
\begin{array}{cc}
g_{k\alpha }^{\downarrow r,}(\varepsilon ) &  \\ 
& g_{k\alpha }^{\uparrow r,a}(\varepsilon )%
\end{array}%
\right) =\left( 
\begin{array}{cc}
\frac{1}{\varepsilon -\varepsilon _{k\alpha \downarrow }+i\eta } &  \\ 
& \frac{1}{\varepsilon -\varepsilon _{k\alpha \uparrow }+i\eta }%
\end{array}%
\right) ,$ and the third-order Green functions defined by: 
\begin{eqnarray*}
G_{kLk^{\prime }\alpha }^{\sigma ^{\prime }\sigma ,1(3)r}(t-t^{\prime })
&=&-i\theta (t-t^{\prime })\langle \{a_{kL\sigma ^{\prime }}(t)a_{k^{\prime
}\alpha \bar{\sigma}^{\prime }}^{\dag }(t)c_{\bar{\sigma}^{\prime
}}(t),c_{\sigma }^{\dag }(t^{\prime })\}\rangle , \\
G_{kLk^{\prime }\alpha }^{\sigma ^{\prime }\sigma ,2(3)r}(t-t^{\prime })
&=&-i\theta (t-t^{\prime })\langle \{a_{kL\sigma ^{\prime }}(t)c_{\bar{\sigma%
}^{\prime }}^{\dag }(t)a_{k^{\prime }\alpha \bar{\sigma}^{\prime
}}(t),c_{\sigma }^{\dag }(t^{\prime })\}\rangle , \\
G_{kLk^{\prime }\alpha }^{\sigma ^{\prime }\sigma ,3(3)r}(t-t^{\prime })
&=&-i\theta (t-t^{\prime })\langle \{a_{kL\bar{\sigma}^{\prime }}^{\dag
}(t)a_{k^{\prime }\alpha \sigma ^{\prime }}(t)c_{\bar{\sigma}^{\prime
}}(t),c_{\sigma }^{\dag }(t^{\prime })\}\rangle , \\
G_{kLk^{\prime }\alpha }^{\sigma ^{\prime }\sigma ,4(3)r}(t-t^{\prime })
&=&-i\theta (t-t^{\prime })\langle \{a_{kL\bar{\sigma}^{\prime }}^{\dag
}(t)c_{\sigma ^{\prime }}(t)a_{k^{\prime }\alpha \bar{\sigma}^{\prime
}}(t),c_{\sigma }^{\dag }(t^{\prime })\}\rangle , \\
G_{kLk^{\prime }\alpha }^{\sigma ^{\prime }\sigma ,5(3)r}(t-t^{\prime })
&=&-i\theta (t-t^{\prime })\langle \{a_{kL\bar{\sigma}^{\prime
}}(t)a_{k^{\prime }\alpha \bar{\sigma}^{\prime }}^{\dag }(t)c_{\sigma
^{\prime }}(t),c_{\sigma }^{\dag }(t^{\prime })\}\rangle , \\
G_{kLk^{\prime }\alpha }^{\sigma ^{\prime }\sigma ,6(3)r}(t-t^{\prime })
&=&-i\theta (t-t^{\prime })\langle \{a_{kL\bar{\sigma}^{\prime }}(t)c_{\bar{%
\sigma}^{\prime }}^{\dag }(t)a_{k^{\prime }\alpha \sigma ^{\prime
}}(t),c_{\sigma }^{\dag }(t^{\prime })\}\rangle .
\end{eqnarray*}%
It is worthily mentioning that one usually takes the Hatree-Fock decoupling
approximation up to the second-order for Green functions in most systems.
Such a decoupling scheme is not adequate for the present single-level QD
coupled system, as it could smear some characteristic features that come
from electronic correlations. In order to extract more information from the
many-body interactions, we should consider the equation of motion up to the
third-order of Green functions $G_{k\alpha }^{i(2)r}(\varepsilon )$\ in Eq. (%
\ref{Self-con}). In above derivations, we have invoked the following
decoupling approximations for Green functions\cite{Haug}: $G_{kLk^{\prime
}\alpha }^{\sigma ^{\prime }\sigma ,1(3)r}(\varepsilon )=G_{kLk^{\prime
}\alpha }^{\sigma ^{\prime }\sigma ,2(3)r}(\varepsilon )=G_{kLk^{\prime
}\alpha }^{\sigma ^{\prime }\sigma ,3(3)r}(\varepsilon )=G_{kLk^{\prime
}\alpha }^{\sigma ^{\prime }\sigma ,6(3)r}(\varepsilon )=0,$ $G_{kLk^{\prime
}\alpha }^{\sigma ^{\prime }\sigma ,4(3)r}(\varepsilon )=-\delta
_{kL,k^{\prime }\alpha }f_{L}(\varepsilon _{kL\bar{\sigma}^{\prime
}})G^{\sigma ^{\prime }\sigma ,r}(\varepsilon ),$ $G_{kLk^{\prime }\alpha
}^{\sigma ^{\prime }\sigma ,5(3)r}(\varepsilon )=\delta _{kL,k^{\prime
}\alpha }[1-f_{L}(\varepsilon _{kL\bar{\sigma}^{\prime }})]G^{\sigma
^{\prime }\sigma ,r}(\varepsilon ).$As we ignore the spin-flip scatterings
in the present system, the off-diagonal elements associated with different
spins of the third-order Green function are sent to zero, namely, the
higher-order spin correlations in the ferromagnetic leads are neglected.

With these results, we may obtain 
\begin{eqnarray*}
G_{kL}^{1(2)r}(\varepsilon ) &=&T_{kL}g_{kL}G^{(2)r}(\varepsilon ), \\
G_{k,q}^{2(2)r}(\varepsilon ) &=&T_{kL}^{\ast }\tilde{g}_{k\alpha
}^{r}[G^{(2)r}(\varepsilon )-\bar{F}_{L}G^{r}(\varepsilon )], \\
G_{k,q}^{3(2)r}(\varepsilon ) &=&-T_{kL}g_{\bar{k}\alpha
}^{r}[G^{(2)r}(\varepsilon )-\bar{F}_{L}G^{r}(\varepsilon )],
\end{eqnarray*}%
with $F_{\alpha }=\left( 
\begin{array}{cc}
f_{\alpha }(\varepsilon _{k\alpha \uparrow }) &  \\ 
& f_{\alpha }(\varepsilon _{k\alpha \downarrow })%
\end{array}%
\right) $ and $\bar{F}_{\alpha }=\left( 
\begin{array}{cc}
f_{\alpha }(\varepsilon _{k\alpha \downarrow }) &  \\ 
& f_{\alpha }(\varepsilon _{k\alpha \uparrow })%
\end{array}%
\right) $.

Similarly, a more straightforward but somewhat complicated calculation gives
rise to the following equations 
\begin{eqnarray}
G_{kR}^{1(2)r}(\varepsilon ) &=&\frac{1}{2}T_{kR}\sin \theta \lbrack
f_{R}(\varepsilon _{kR\uparrow })g_{kR}^{\uparrow r}(\varepsilon
)-f_{R}(\varepsilon _{kR\downarrow })g_{kR}^{\downarrow r}(\varepsilon )]%
\hat{\sigma}_{1}G^{r}(\varepsilon )  \notag \\
&&+T_{kR}\left[ \cos ^{2}\frac{\theta }{2}g_{kR}^{r}(\varepsilon )+\sin ^{2}%
\frac{\theta }{2}g_{\bar{k}R}^{r}(\varepsilon )\right] G_{q,q}^{(2)r}(%
\varepsilon )
\end{eqnarray}%
\begin{eqnarray}
G_{kR}^{2(2)r}(\varepsilon ) &=&\frac{1}{2}T_{kR}\sin \theta \lbrack \tilde{g%
}_{kR}^{\downarrow ,r}(\varepsilon )f_{R}(\varepsilon _{kR\uparrow })-\tilde{%
g}_{kR}^{\uparrow ,r}(\varepsilon )f_{R}(\varepsilon _{kR\downarrow })]\hat{%
\sigma}_{1}G^{r}(\varepsilon )  \notag \\
&&-T_{kR}\left[ \cos ^{2}\frac{\theta }{2}\tilde{g}_{kR}^{r}(\varepsilon )%
\bar{F}_{R}+\sin ^{2}\frac{\theta }{2}\tilde{g}_{\bar{k}R}^{r}(\varepsilon
)F_{R}\right] G^{r}(\varepsilon )  \notag \\
&&+\frac{1}{2}T_{kR}\sin \theta \lbrack \tilde{g}_{kR}^{\uparrow
,r}(\varepsilon )-\tilde{g}_{kR}^{\downarrow ,r}(\varepsilon )]\hat{\sigma}%
_{1}G^{(2)r}(\varepsilon )  \notag \\
&&+T_{kR}\left[ \cos ^{2}\frac{\theta }{2}\tilde{g}_{kR}^{r}(\varepsilon
)+\sin ^{2}\frac{\theta }{2}\tilde{g}_{\bar{k}R}^{r}(\varepsilon )\right]
G^{(2)r}(\varepsilon ),
\end{eqnarray}%
\begin{eqnarray}
G_{kR}^{3(2)r}(\varepsilon ) &=&-T_{kR}^{\ast }\left[ \sin ^{2}\frac{\theta 
}{2}g_{kR}^{r}(\varepsilon )+\cos ^{2}\frac{\theta }{2}g_{\bar{k}%
R}^{r}(\varepsilon )\right] G^{(2)r}(\varepsilon )  \notag \\
&&+T_{kR}^{\ast }\left[ \sin ^{2}\frac{\theta }{2}g_{kR}^{r}(\varepsilon
)F_{R}+\cos ^{2}\frac{\theta }{2}g_{\bar{k}R}^{r}(\varepsilon )\bar{F}_{R}%
\right] G^{r}(\varepsilon ),
\end{eqnarray}%
where $\hat{\sigma}_{1}=\left( 
\begin{array}{cc}
0 & 1 \\ 
1 & 0%
\end{array}%
\right) ,$ and $\tilde{g}_{\bar{k}\alpha }^{r}(\varepsilon )=\left( 
\begin{array}{cc}
\tilde{g}_{k\alpha }^{\downarrow r}(\varepsilon ) &  \\ 
& \tilde{g}_{k\alpha }^{\uparrow r}(\varepsilon )%
\end{array}%
\right) =\left( 
\begin{array}{cc}
\frac{1}{\varepsilon -2\varepsilon _{0}-U+\varepsilon _{k\alpha \uparrow
}+i\eta } &  \\ 
& \frac{1}{\varepsilon -2\varepsilon _{0}-U+\varepsilon _{k\alpha \downarrow
}+i\eta }%
\end{array}%
\right) $.

On the other hand, the second-order retarded Green's function can be
expressed as of the following form$\ \ \ $%
\begin{equation}
W_{1}G^{(2)r}(\varepsilon )=N+W_{2}G^{r}(\varepsilon ),  \label{3rd-order}
\end{equation}%
where 
\begin{eqnarray*}
W_{1} &=&(\varepsilon -\varepsilon _{0}-U)-\underset{kL}{\sum }%
T_{kL}T_{kL}^{\ast }g_{kL}^{r}(\varepsilon )-\underset{kL}{\sum }%
T_{kL}T_{kL}^{\ast }\tilde{g}_{kL}^{r}(\varepsilon )-\underset{kL}{\sum }%
T_{kL}T_{kL}^{\ast }g_{\bar{k}L}^{r}(\varepsilon ) \\
&&\underset{kR}{-\sum }T_{kR}T_{kR}^{\ast }\left[ \cos ^{2}\frac{\theta }{2}%
g_{kR}^{r}(\varepsilon )+\sin ^{2}\frac{\theta }{2}g_{\bar{k}%
R}^{r}(\varepsilon )\right] \\
&&-\underset{kR}{\sum }T_{kR}T_{kR}^{\ast }\frac{1}{2}\sin \theta \left[ 
\tilde{g}_{kR}^{\uparrow ,r}(\varepsilon )-\tilde{g}_{kR}^{\downarrow
,r}(\varepsilon )\right] \hat{\sigma}_{1} \\
&&-\underset{kR}{\sum }T_{kR}T_{kR}^{\ast }\left[ \cos ^{2}\frac{\theta }{2}%
\tilde{g}_{kR}^{r}(\varepsilon )+\sin ^{2}\frac{\theta }{2}\tilde{g}_{\bar{k}%
R}^{r}(\varepsilon )\right] \\
&&-\underset{kR}{\sum }T_{kR}T_{kR}^{\ast }\left[ \cos ^{2}\frac{\theta }{2}%
g_{\bar{k}R}^{r}(\varepsilon )+\sin ^{2}\frac{\theta }{2}g_{\bar{k}%
R}^{r}(\varepsilon )\right] ,
\end{eqnarray*}

\begin{eqnarray*}
W_{2} &=&-\underset{kL}{\sum }T_{kL}T_{kL}^{\ast }\tilde{g}%
_{kL}^{r}(\varepsilon )\bar{F}_{L}-\underset{kL}{\sum }T_{kL}T_{kL}^{\ast
}g_{\bar{k}L}^{r}(\varepsilon )\bar{F}_{L} \\
&&+\underset{kR}{\sum }T_{kR}T_{kR}^{\ast }\frac{1}{2}\sin \theta \lbrack
g_{kR}^{\uparrow ,r}(\varepsilon )f_{R}(\varepsilon _{kR\uparrow
})-g_{kR}^{\downarrow ,r}(\varepsilon )f_{R}(\varepsilon _{kR\downarrow })]%
\hat{\sigma}_{1} \\
&&+\underset{kR}{\sum }T_{kR}T_{kR}^{\ast }\frac{1}{2}\sin \theta \lbrack 
\tilde{g}_{kR}^{\downarrow ,r}(\varepsilon )f_{R}(\varepsilon _{kR\uparrow
})-\tilde{g}_{kR}^{\uparrow ,r}(\varepsilon )f_{R}(\varepsilon
_{kR\downarrow })]\hat{\sigma}_{1} \\
&&-\underset{kR}{\sum }T_{kR}T_{kR}^{\ast }\left[ \cos ^{2}\frac{\theta }{2}%
\tilde{g}_{kR}^{r}(\varepsilon )\bar{F}_{R}+\sin ^{2}\frac{\theta }{2}\tilde{%
g}_{\bar{k}R}^{r}(\varepsilon )F_{R}\right] \\
&&-\underset{kR}{\sum }T_{kR}T_{kR}^{\ast }\left[ \cos ^{2}\frac{\theta }{2}%
g_{\bar{k}R}^{r}(\varepsilon )\bar{F}_{R}+\sin ^{2}\frac{\theta }{2}%
g_{kR}^{r}(\varepsilon )F_{R}\right] .
\end{eqnarray*}%
By combining Eq. (\ref{3rd-order}) with Eq. (\ref{Self-con}), we get

\begin{equation}
(\varepsilon -\varepsilon _{0}-\Sigma _{0}^{r}-UW_{1}^{-1}W_{2})G^{r}=%
\mathbf{1}+UW_{1}^{-1}N.  \label{eqn-motion-f}
\end{equation}%
From this equation, the retarded Green function $G^{r}$ can be obtained. We
would like to point out that the averaged values involved in Eq. (\ref%
{Self-con}) should be obtained self-consistently by%
\begin{eqnarray*}
\langle n_{\sigma }\rangle &=&\Im m\int \frac{d\varepsilon }{2\pi }G^{\sigma
\sigma <}(\varepsilon )\text{,} \\
\langle c_{\sigma }^{\dag }c_{\bar{\sigma}}\rangle &=&-i\int \frac{%
d\varepsilon }{2\pi }G^{\bar{\sigma}\sigma <}(\varepsilon ).
\end{eqnarray*}%
These above-mentioned equations establish the fundamental basis for
numerically investigating the spin-dependent transport properties of the
FM-QD-FM coupled MTJs.

\section{Numerical Results}

Without losing the generality, in the following numerical calculations we
may further suppose that the two ferromagnets are made of the same
materials, i.e., $P_{L}=P_{R}=P$, where $P_{L(R)}=(\Gamma _{L(R)\uparrow
}-\Gamma _{L(R)\downarrow })/(\Gamma _{L(R)\uparrow }+\Gamma
_{L(R)\downarrow })$ is the polarization of the left (right) ferromagnet.
Then, we can introduce $\Gamma _{L\uparrow ,\downarrow }=\Gamma _{R\uparrow
,\downarrow }=\Gamma _{0}(1\pm P)$, where $\Gamma _{0}=$ $\Gamma
_{L(R)\uparrow }(P=0)=\Gamma _{L(R)\downarrow }(P=0)$ will be taken as an
energy scale. The tunnel matrix elements $T_{kL(R)}$ is presumed to take
values at the Fermi level. We will take $I_{0}=\frac{e\Gamma _{0}}{\hbar }$
and $G_{0}=\frac{e^{2}}{\hbar }$ as scales for the currents and
corresponding differential conductance, respectively. 
\begin{figure}[tbp]
\includegraphics[width=9cm,height=9cm,clip]{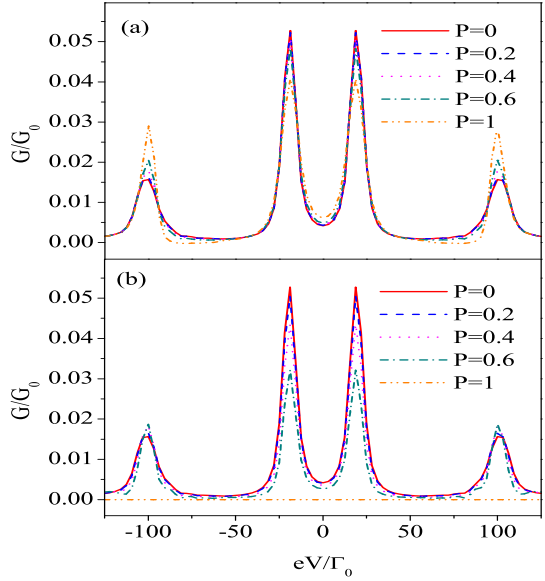} \vspace{-0.5cm}
\caption{The bias dependence of the differential conductance $G$ for
different polarization $P$ in parallel (a) and antiparallel (b)
configurations of magnetizations. The parameters are taken as $\protect%
\varepsilon _{0}=10\Gamma _{0}$, $U=40\Gamma _{0}$, $k_{B}T=0.3\Gamma _{0}$.}
\label{Fig. 2}
\end{figure}

\subsection{Differential conductance}

The bias dependences of the tunnel electrical current and the differential
conductance $G=\frac{dI}{dV}$ in the FM-QD-FM coupled system have been
reexamined within the present scheme. It is found that the tunnel electrical
current exhibits two-step features, which corresponds to the resonant
tunneling of electrons through the QD at energy levels $\varepsilon _{0}$\
and $\varepsilon _{0}+U$. The corresponding differential conductance as a
function of the bias thus exhibits resonant peaks, where the large peak at a
lower voltage corresponds to the discrete level $\varepsilon _{0}$, and a
small resonant peak corresponds to the level $\varepsilon _{0}+U$. It is
uncovered that with the increase of the interaction $U$, the peaks of $G(V)$%
\ corresponding to the discrete level $\varepsilon _{0}+U$\ shift towards
the higher voltage side without changing the peak amplitude for both
parallel and antiparallel configurations. This is because an increase of $U$%
\ makes two discrete levels $\varepsilon _{0}$\ and $\varepsilon _{0}+U$\
become far apart, thereby leading to that the resonant positions of main and
charging peaks of $G(V)$ are far\textit{\ }separated. These observations are
quite consistent with the previous studies (see e.g. Ref.\cite{Rudzinski}).

The bias dependence of the differential conductance for different
polarizations in the parallel and antiparallel configurations of
magnetization is shown in Fig. 2. It can be found that in the parallel
configuration there is only slight changes for different polarizations
except for $P=1$, as displayed in Fig. 2(a). We can understand this property
from the point of view of the resonant tunneling through two barriers\cite%
{Sergueev}, namely, the two barriers are of the same height, and the
resonant probability is unity in this case, that is independent with $P$.
The small changes with $P$\ observed in Fig. 2(a) is caused by the
interaction $U$. However, it is quite different for the case with full
polarization $P=1$, which comes from the fact that the conductance for $P=1$%
\ has only one channel. It is interesting to note that with increasing $P$\
in the case of parallel alignment, although the main peaks at the bias
corresponding to the energy level $\varepsilon _{0}$\ are suppressed, the
charging peaks at the bias corresponding to the level $\varepsilon _{0}+U$\
are enhanced. In this case, more electrons from the left enter into the QD
through the channel $\varepsilon _{0}+U$\ to tunnel into the right lead.\
For the antiparallel configuration, it can be seen that the main peaks of
the conductance decrease as the polarization $P$\ increases, while the
charging peaks remain almost intact, as shown in Fig. 2(b), that is
different from the case of parallel configuration, and is due to the
conventional spin-valve effect in MTJs. For $P=1$, the conductance becomes
zero, which is nothing but a perfect spin-valve effect.

\begin{figure}[tbp]
\includegraphics[width=9cm,height=9cm,clip]{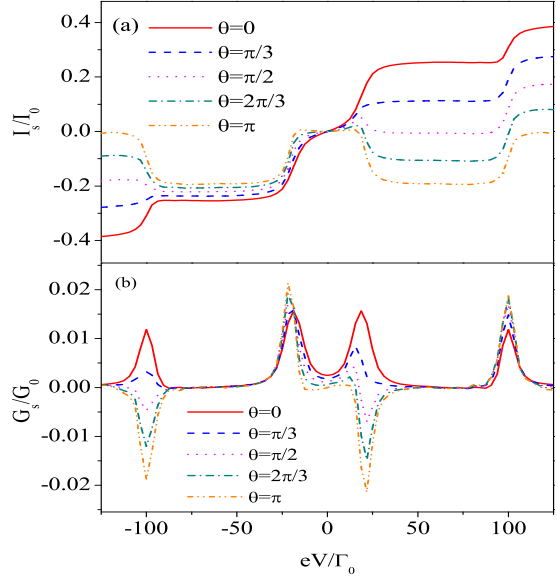} \vspace{-0.5cm}
\caption{The bias dependence of the spin current $I_{s}$ (a) and spin
differential conductance $G_{s}$ (b) for different angle $\protect\theta $
at $P=0.4$. The other parameters are taken the same as in Fig. 2.}
\label{Fig. 3}
\end{figure}

\subsection{Spin current and spin differential conductance}

The bias dependence of the spin current $I_{s}$ and its corresponding spin
differential conductance defined by $G_{s}=\frac{dI_{s}}{dV}$ for different
angles is shown in Fig. 3. It can be seen that, by changing the relative
angle $\theta $, the spin current can change its magnitude even its sign. At 
$V>0$, when $\theta <\pi /2$, the spin current $I_{s}$ shows step-like
behaviors, and $I_{s}>0$; while for $\theta >\pi /2$, $I_{s}$ exhibits
behaviors similar to a basin-like shape, and $I_{s}<0$ at the bottom of the
"basin". This shows that the effect of the relative orientation of
magnetizations on the spin current is more obvious. At two resonant
positions corresponding to levels $\varepsilon _{0}$ and $\varepsilon _{0}+U$%
, $I_{s}$ begins to sharply change. It is clearly seen in Fig. 3(b) that at $%
V>0$, the spin conductance shows sharp resonant peaks and dips at the levels 
$\varepsilon _{0}$ and $\varepsilon _{0}+U$, although for $\theta =0$ and $%
\pi /3$ the resonance corresponding to $\varepsilon _{0}$ shows a peak,
while for $\theta =\pi /2$, $2\pi /3$ and $\pi $ the first resonance shows a
sharp dip. The resonances corresponding to the level $\varepsilon _{0}+U$
exhibit peaks for all $\theta $. At $\theta =0$, the two resonant peaks of $%
I_{s}$ have comparable weights; at $\theta =\pi /3$, the first peak has a
small weight, while the second peak has a large weight; at $\theta =\pi /2$,
the first dip of $I_{s}$ has a small amplitude, while the other peak has a
large amplitude; at $\theta =2\pi /3$, the resonant dip and peak have almost
the same amplitude; and at $\theta =\pi $, the first dip has a larger
amplitude than the peak. Evidently, the spin current and spin conductance
have no spin-valve effect, showing that the spin current and spin
conductance have quite different characteristics from their electrical
counterparts.

\begin{figure}[tbp]
\includegraphics[width=9cm,height=9cm,clip]{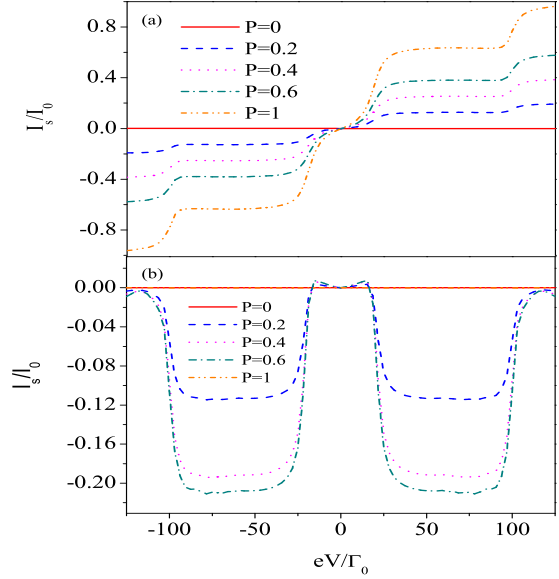} \vspace{-0.5cm}
\caption{The bias dependence of the spin current $I_{s}$ for different
polarization $P$ in parallel (a) and antiparallel (b) configurations of
magnetizations. The other parameters are taken the same as in Fig. 2.}
\label{Fig. 4}
\end{figure}

The step-like and basin-like behaviors of the bias dependence of the spin
current for different polarizations in parallel and antiparallel
configurations are presented in Figs. 4(a) and (b), respectively. For $P=0$,
as there is no polarization in electrodes, the contribution of electrons
with spin up is the same as that of spin down, leading to the spin current
vanishes. In the case of parallel alignment ($\theta =0$), with increasing
the polarization $P$, the contribution of spin up is increased while that of
spin down is decreased, thus resulting in the magnitude of the spin current
increased, as shown in Fig. 4(a). The step-like behavior of $I_{s}$ against $%
V$ comes from the resonant tunneling. The situation becomes a bit
complicated in the case of antiparallel alignment ($\theta =\pi $). When $%
P=0 $ and $1$, the spin current is zero, because for the former $I_{\uparrow
}=I_{\downarrow }$ while for the latter $I_{\uparrow }=I_{\downarrow }=0$,
as shown in Fig. 4(b). For $P\neq 0$ and $1$, with increasing the bias
voltage the spin current decreases steeply at the resonant position
corresponding to the QD energy level $\varepsilon _{0}$, and then keeps a
constant until the other resonant position corresponding to the level $%
\varepsilon _{0}+U$ where $I_{s}$ increases sharply, leading to the spin
current shows a basin-like behavior. The larger the polarization $P$ $(\neq
0,1)$, the deeper the bottom of the basin. These results demonstrate that
the spin current $I_{s}$ has quite different behaviors in parallel and
antiparallel configurations.

\begin{figure}[tbp]
\includegraphics[width=9cm,height=9cm,clip]{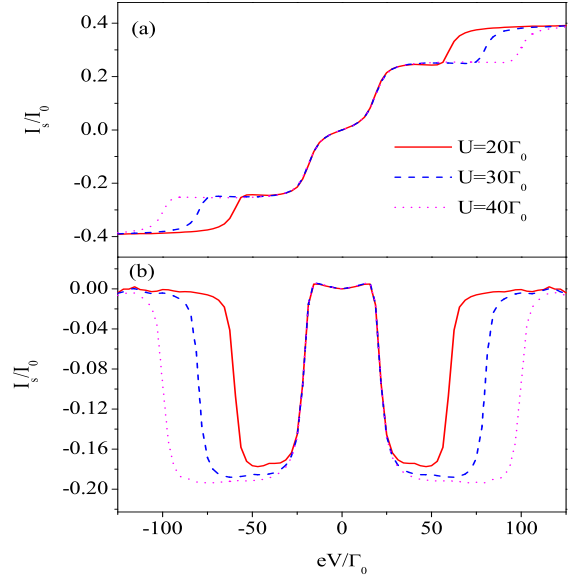} \vspace{-0.5cm}
\caption{The bias dependence of spin current $I_{s}$ in parallel (a) and
antiparallel (b) configurations for different Coulomb interaction $U$, where 
$P=0.4$, and the other parameters are taken the same as in Fig. 2.}
\label{Fig. 5}
\end{figure}

Fig. 5 shows the bias dependence of spin current in collinear configurations
for different Coulomb interaction $U$. It can be found that the overall
qualitative behavior of $I_{s}$ against $V$ looks similar to that shown in
Fig. 4. With increasing $U$, in the parallel alignment the second plateau of
the spin current $I_{s}$ becomes wider, as shown in Fig. 5(a); while in the
antiparallel alignment apart from that the bottom of the "basin" becomes
deeper, the width of the "basin" also becomes wider, as shown in Fig. 5(b).
This is again from the fact that an increase of $U$ makes the resonant
positions corresponding QD discrete levels $\varepsilon _{0}$ and $%
\varepsilon _{0}+U$ become far separated.

\begin{figure}[tbp]
\includegraphics[width=9cm,height=9cm,clip]{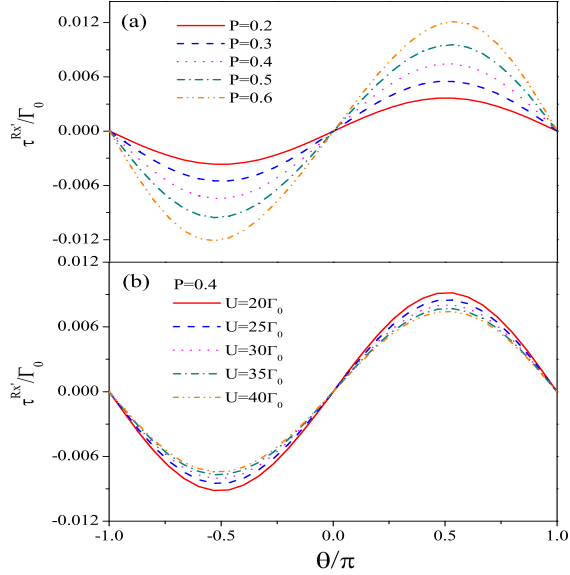} \vspace{-0.5cm}
\caption{The current-induced spin transfer torque against the relative
orientation angle $\protect\theta $ for different polarization $P$ (a) and
different interaction $U$ (b) at the bias voltage $V=25\Gamma _{0}/e$. The
other parameters are taken the same as in Fig. 2.}
\label{Fig. 6}
\end{figure}

\subsection{Current-induced spin transfer torque}

The angular dependence of the CISTT for different polarization and Coulomb
interaction is shown in Fig. 6. It is seen that the CISTT versus the angle $%
\theta $ shows a sine-like behavior, consistent with the finding in Ref.\cite%
{Slonc}. At $\theta =0$ and $\pi $, i.e., the magnetizations of the two
ferromagnetic electrodes are collinearly aligned, the CISTT vanishes, which
is obvious as the spin torque $\propto \mathbf{S}_{R}\times (\mathbf{S}%
_{L}\times \mathbf{S}_{R})$, where $\mathbf{S}_{L}$ and $\mathbf{S}_{R}$ are
the spin moments of the left and right ferromagnets. With increasing the
polarization $P$, the magnitude of the CISTT is enhanced, as displayed in
Fig. 6(a). This is in agreement with the statement that the torque is
proportional to the polarization of the other ferromagnet\cite{Slonc}. With
the increase of Coulomb interaction $U$, the magnitude of the CISTT is
decreased, as seen in Fig. 6(b). This can be understood in such a way that
the contribution of the tunneling through the discrete level $\varepsilon
_{0}+U$ is decreased with increasing the interaction $U$, thus leading to a
decrease of the spin transfer torque.

\begin{figure}[tbp]
\includegraphics[width=9cm,height=9cm,clip]{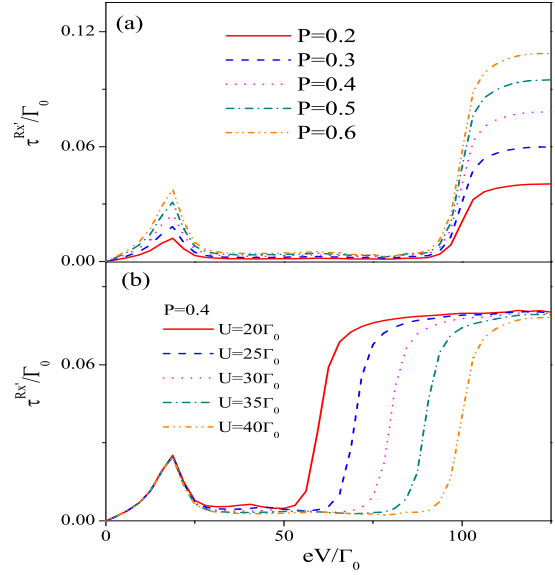} \vspace{-0.5cm}
\caption{The bias dependence of the current-induced spin transfer torque for
different polarization $P$ (a) and $U$ (b) at $\protect\theta =$ $\protect%
\pi /3$. The other parameters are taken the same as in Fig. 2.}
\label{Fig. 7}
\end{figure}

The bias dependences of the CISTT for different polarizations and Coulomb
interactions at a given angle $\theta =\pi /3$ are shown in Figs. 7(a) and
(b). It is seen that with increasing the bias voltage, the CISTT first
increases slowly to a sharp peak, then decreases dramatically to almost
zero, and after undergoing a wider flat, it increases suddenly to another
plateau. The position of the first sharp peak in the curve of CISTT versus $%
V $ is observed to be independent of the polarization as well as the Coulomb
interaction, showing that it is a resonant peak at the resonant position $%
\varepsilon _{0}$. The second sharp increase of the CISTT against $V$ is
also from the resonance at the QD discrete energy level $\varepsilon _{0}+U$%
. From Fig. 7(a), one may find that with increasing the polarization the
magnitudes of the CISTT at the two resonant positions increase. This is
because the transmission coefficient of electrons is proportional to the
polarization\cite{zhu1}, while the CISTT depends on the tunneling electrical
current that is determined by the transmission coefficient.\textit{\ }Thus,
an increase of polarization would enhance the CISTT at the resonant
positions. The effect of the Coulomb interaction $U$ on the CISTT is shown
in Fig. 7(b). At the first peak, it is independent of $U$, suggesting that
it is a resonance at the level $\varepsilon _{0}$; while with increasing $U$%
, the second resonant positions move to higher voltages, showing that the
second resonance takes place at the level $\varepsilon _{0}+U$. It is
interesting to note that at a given polarization the CISTT goes to a
saturation plateau at the bias larger than $(\varepsilon _{0}+U)/e$, which
is independent of the Coulomb interaction. From Fig. 7, we could see that
the CISTT can be remarkably enhanced at the QD discrete energy levels which
may be adjusted by changing the gate voltage.

\begin{figure}[tbp]
\includegraphics[width=9cm,height=9cm,clip]{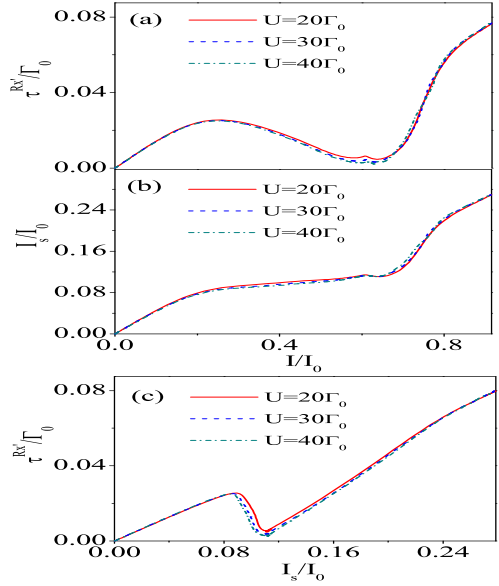} 
\caption{The current-induced spin transfer torque (a) and the spin current
(b) against the electrical current as well as the current-induced spin
transfer torque versus the spin current (c) for different $U$ at $\protect%
\theta =$ $\protect\pi /3$ and $P=0.4$. The other parameters are taken the
same as in Fig. 2.}
\label{Fig. 8}
\end{figure}

What is the relationship between the CISTT, the electrical current and the
spin current in this FM-QD-FM coupled system? The answer is presented in
Fig. 8 for different $U$ at $\theta =$ $\pi /3$. With increasing the
electrical current, the CISTT goes up linearly to a round peak, then
decreases slowly to a broad minimum, and then increases sharply, as shown in
Fig. 8(a). The first round peak is owing to the resonant tunneling at the
level $\varepsilon _{0}$; while the second sharp increase is caused by the
resonant tunneling at the level $\varepsilon _{0}+U$. The spin current $%
I_{s} $ is found to increase non-monotonically with increasing the
electrical current $I$, as illustrated in Fig. 8(b). This is quite different
from the spin current in FM-SC-FM tunnel junctions where the spin current is
proportional to the injection current\cite{Takahashi}. In the curves of $%
I_{s}$ against $I$, the two resonances can be clearly seen. The CISTT as a
function of $I_{s}$ shows a kink-like behavior, as displayed in Fig. 8(c).
With increasing $I_{s}$, the CISTT first shows a linear behavior, then
suddenly drops to nearly zero, and then increases almost linearly again.
This kink-like behavior is caused by the resonant tunneling at the levels $%
\varepsilon _{0}$ and $\varepsilon _{0}+U$, which manifests the same thing
as in Fig. 7.\textit{\ }Away from the kink, the CISTT appears to be
proportional to the spin current. The reason is that the electrons with
opposite spins from the left ferromagnetic electrode entering into the right
electrode exert the torques with opposite directions on the spins of the
right electrode, giving rise to that the magnitude of the CISTT is
determined by the difference between the magnitudes of torques exerted by
electrons with spin up and spin down. Considering that the spin current is
defined by the difference between the electrical currents of spin up and
spin down, while the spin transfer torque is proportional to the electrical
current\cite{Slonczewski,Berger}, the proportionality between the CISTT and
the spin current is conceivable.\textit{\ }It is seen that for different $U$
the curves in Fig. 8 are almost unchanged, showing that the Coulomb
interaction has less effect on the CISTT versus $I$ and $I_{s}$, as well as $%
I_{s}$ versus $I$. This is because the Coulomb interaction $U$ only changes
the resonant positions.

\section{Summary}

We have investigated the spin-dependent transport properties in a FM-QD-FM
coupled system by means of the nonequilibrium Green functions. It has been
found that by changing the relative orientation of both magnetizations, the
spin current can change its magnitude even its sign. For positive bias
voltages, when the relative orientation angle is less than $\pi /2$, the
spin current shows a step-like behavior, and the spin current is positive;
when the relative orientation angle is greater than $\pi /2$, the spin
current behaviors similar to a basin-like shape, and is negative at the
bottom of the "basin". With increasing the Coulomb interaction $U$ in the
QD, it has been uncovered that the resonant positions for the tunneling
electrical current and spin current become far separated. The CISTT is
observed to first increase slowly to a sharp peak with increasing the bias
voltage, then decrease dramatically to almost zero; and after undergoing an
unchanged stage, it increases suddenly to another plateau. Such a behavior
is obviously resulted from the resonant tunneling through the central QD.
The magnitudes of the CISTT at two resonant positions are observed to
increase with increasing the polarization of the ferromagnet. At the bias
larger than $(\varepsilon _{0}+U)/e$, the CISTT is seen to reach a
saturation plateau which is independent of the Coulomb interaction. In
addition, it has been demonstrated that the CISTT as a function of the spin
current shows a kink-like behavior, and away from the kink, the CISTT
increases almost linearly with increasing the spin current. The CISTT as a
function of the electrical current shows a non-monotonic behavior, while the
spin current is found to be nonlinearly proportional to the electrical
current. Besides, the Coulomb interaction is shown to have less effect on
the behaviors of the CISTT as functions of the tunneling electrical current
and spin current, as well as the spin current as a function of the
electrical current.

Finally, we would like to mention that the CISTT can be greatly enhanced
when the bias voltage meets with the discrete levels of the QD at two
resonant positions. While the energy level of the QD can be adjusted by
tuning the gate voltage, the resonant property of the CISTT offers an
alternative premise to develop a spintronic device through the
current-controlled magnetization reversal effect.

\acknowledgments We have benefitted from discussions with B. Jin, Z. C.
Wang, and Z. G. Zhu. This work is supported in part by the National Science
Foundation of China (Grants Nos. 90403036, 20490210, 10247002) and by the
Chinese Academy of Sciences.

\end{document}